\documentclass{pasa}%

\usepackage{graphicx}
\usepackage{siunitx}
\usepackage{amsmath}
\usepackage{mathtools}
\newcommand{\sdots}{\raisebox{3pt}{$\scalebox{.75}{\dots}$}}
\newcommand{\svdots}{\raisebox{3pt}{$\scalebox{.75}{\vdots}$}}
\newcommand{\sddots}{\raisebox{3pt}{$\scalebox{.75}{$\ddots$}$}}

\usepackage{float}
\usepackage{tabularx}

\title[Peeling with Shapelets]{Modelling and peeling extended sources with shapelets: a Fornax~A case study}

\author[J.~L.~B.~Line et al.]{J.~L.~B.~Line$^{1,2}$\thanks{jack.line@curtin.edu.au}, 
D.~A.~Mitchell$^3$,
B.~Pindor$^{4,2}$,
J.~L.~Riding$^{4,2}$,
B.~McKinley$^{1,2}$,
R.~L.~Webster$^{4,2}$,
C.~M.~Trott$^{1,2}$,
N.~Hurley-Walker$^{1,2}$,
A.~R.~Offringa$^5$
\\
\affil{$^1$International Centre for Radio Astronomy Research, Curtin University, Perth, WA 6845, Australia}
\affil{$^2$ARC Centre of Excellence for All Sky Astrophysics in 3 Dimensions (ASTRO-3D)}
\affil{$^3$CSIRO Astronomy and Space Science (CASS), PO Box 76, Epping, NSW 1710, Australia}
\affil{$^4$The University of Melbourne, School of Physics, Parkville, VIC 3010, Australia}
\affil{$^5$Netherlands Institute for Radio Astronomy (ASTRON), 7991 PD Dwingeloo, The Netherlands}
}%

\jid{PASA}
\doi{10.1017/pas.\the\year.xxx}
\jyear{\the\year}

\usepackage{aas_macros}
\usepackage{hyperref} 
\hypersetup{colorlinks,citecolor=blue,linkcolor=blue,urlcolor=blue}

\begin{document}
\sisetup{tight-spacing=true}

\begin{frontmatter}
\maketitle

\begin{abstract}
To make a power spectrum (PS) detection of the 21$\,$cm signal from the Epoch of Reionisation (EoR), one must avoid/subtract bright foreground sources. Sources such as Fornax~A present a modelling challenge due to spatial structures spanning from arc seconds up to a degree. We compare modelling with multi-scale (\texttt{MS}) \texttt{CLEAN} components to `shapelets', an alternative set of basis functions. We introduce a new image-based shapelet modelling package, \texttt{SHAMFI}. We also introduce a new \texttt{CUDA} simulation code (\texttt{WODEN}) to generate point source, Gaussian, and shapelet components into visibilities. We test performance by modelling a simulation of Fornax A, peeling the model from simulated visibilities, and producing a residual PS. We find the shapelet method consistently subtracts large-angular-scale emission well, even when the angular-resolution of the data is changed. We find that when increasing the angular-resolution of the data, the \texttt{MS CLEAN} model worsens at large angular-scales. When testing on real MWA data, the expected improvement is not seen \textit{in real data} because of the other dominating systematics still present. Through further simulation we find the expected differences to be lower than obtainable through current processing pipelines. We conclude shapelets are worthwhile for subtracting extended galaxies, and may prove essential for an EoR detection in the future, once other systematics have been addressed.
\end{abstract}

\end{frontmatter}

\section{Introduction}
\label{sec:intro}

Detecting primordial hydrogen during the Epoch of Reionisation via the 21$\,$cm line power spectrum has been a hotly pursued goal in astrophysics over the last decade, with projects on established instruments such as LOFAR~\citep{VanHaarlem2013}, MWA~\citep{Tingay2013}, and PAPER~\citep{Parsons2010}, and future projects on HERA~\citep{DeBoer2016} and SKA\_{}LOW~\citep{Dewdney2013} all pushing for a detection. The signal promises to constrain cosmological models and fundamental physics~\citep[see][for reviews]{Morales2009,Pritchard2012,Mellema2013}, but is swamped by intervening emission from other astrophysical objects, including radio-loud galaxies and diffuse synchrotron emission. These foregrounds can be up to $\sim5$ orders of magnitude brighter than the 21$\,$cm signal. Recent work and power spectrum limits from LOFAR~\citep{Patil2017,Mertens2020}, MWA~\citep{Barry2019,Li2019,Trott2020}, and PAPER~\citep{Kerrigan2018} all point towards painstakingly precise foreground subtraction being needed to facilitate a detection, with a strong focus on spectral smoothness. As these instruments are interferometers, which make measurements in the Fourier transform of image space called `visibilities', any foreground models benefit from easily being calculated in both image and visibility space.

Some radio-loud galaxies are nearby and cover a large angular extent on the sky (up to degrees in angle), and display rich morphologies, e.g. Centaurus A~\citep[]{McKinley2013}. These sources present a modelling challenge, and are bright enough to necessitate careful subtraction when present in EoR data. The most straight-forward prescriptions build extended models out of Dirac delta functions, which also lends synergy with \texttt{CLEAN}-based deconvolution imaging techniques. For the most extended objects this requires a large number of components, which can become computationally demanding. Furthermore, as discussed in \citet{Yatawatta2010}, the choice of pixel size in image deconvolution for extended emission affects the modelled emission.

An alternative modelling approach is to use so-called `shapelets', a set of orthonormal basis functions consisting of weighted Hermite polynomials~\citep{Refregier2003}. These basis functions are attractive due to having analytically-defined Fourier transforms, and so can be fitted in image space and then generated directly in visibility space. They have been used in a number of novel astronomical applications including: modelling three-dimensional distribution of dust~\citep{Schechtman-Rook2012}; weak lensing measurements in simulations of radio images~\citep{Bacon2014}; gravitationally lensed images~\citep{Tagore2016}; classifying bent radio galaxies~\citep{Bastien2017}. Furthermore, they can be scaled to be extended on the sky, and lend themselves to compression/truncation. \citet{Yatawatta2010} replaced \texttt{CLEAN} components with two dimensional shapelet basis functions around Cygnus A during image-based deconvolution, and was able to increase the dynamic range in an image of Westerbork Synthesis Radio Telescope data by a factor of $\sim$50. Calibration packages such as \texttt{SAGECAL}\footnote{\url{https://github.com/nlesc-dirac/sagecal}}~\citep{Yatawatta2009,Kazemi2010} and the \texttt{RTS}~\citep{Mitchell2008,Riding2017} have the capability to use shapelet models, with \texttt{SAGECAL} shapelet models being used predominantly with LOFAR data, as described in the LOFAR cookbook\footnote{\url{https://support.astron.nl/LOFARImagingCookbook/shapelets.html}}.

Recent advances in \texttt{CLEAN} algorithms and packages now allow multiscale (\texttt{MS}) \texttt{CLEAN}ing~\citep[see][and references therein]{Cornwell2008b}, with the option to use Gaussians during the \texttt{CLEAN}, as well as point sources\footnote{We note that as unresolved sources, known as point sources, are modelled using Dirac delta functions, when we refer to a point source in a simulation, we are referring to modelling a component using a Dirac delta function}. \texttt{WSClean}\footnote{\url{https://sourceforge.net/projects/wsclean/}}~\citep{Offringa2014,Offringa2017} is a cutting-edge imaging package capable of all the above, and has the ability to produce point source and Gaussian models. The main goal of this paper is to compare a \texttt{MS CLEAN} component model to a shapelet model, testing how effective and computationally efficient they are during calibration and peeling, with the primary metric being the 2D 21$\,$cm power spectrum as used by EoR experiments. We chose to use Fornax~A as a case-study, as it sits in the primary beam side-lobes of one of the MWA EoR observational fields~\citep[see][for more details on MWA EoR processing pipelines]{Jacobs2016}, and as such must be peeled from the data. 

In this paper we introduce a new image-based shapelet fitting package, \texttt{SHA}pelet \texttt{M}odelling \texttt{F}or \texttt{I}nterferometers (\texttt{SHAMFI}\footnote{\url{https://github.com/JLBLine/SHAMFI}}, Section~\ref{subsec:fit_with_shamfi}), which we use to test against \texttt{MS CLEAN} outputs from \texttt{WSClean}. To test the efficiency of generating each type of model, as well as their ability to accurately subtract EoR foreground emission, we have also developed the visibility simulator \texttt{WODEN}\footnote{\url{https://github.com/JLBLine/WODEN.git} \texttt{WODEN} is not an acronym, it's named after the Anglo-Saxon pagan god} (see Section~\ref{sec:woden}). We first simulate data using \texttt{WODEN} to test each method in the absence of instrumental, atmospheric, and astrophysical contaminants. We then test each method on real MWA data using the~\texttt{RTS}. 

The paper is organised as follows. In Section~\ref{sec:shape_fitting} we introduce shapelets as basis functions, and introduce \texttt{SHAMFI}. In Section~\ref{sec:woden} we introduce the GPU-accelerated visibility simulator \texttt{WODEN}. In Section~\ref{sec:sim_forA} we introduce a model from which to simulate Fornax~A data, outline simulations of this model with \texttt{WODEN}, and detail fitting the simulations using \texttt{SHAMFI}. In Section~\ref{sec:sim_peel-results} we introduce a method to peel with the simulated data, and present simulated peeling results, including testing truncation on both the shapelet and \texttt{MS CLEAN} models. In Section~\ref{sec:real_data} we fit and peel using real data. We discuss our results in Section~\ref{sec:discuss}, and summarise in Section~\ref{sec:summarise}.

\section{Shapelet fitting and SHAMFI}
\label{sec:shape_fitting}
Shapelets as a basis function are introduced in~\citet{Refregier2003}, to which we refer the reader for a detailed overview; we briefly define them here for clarity within this paper. Shapelets are orthonormal basis functions based around hermite polynomials, and can be based in a polar or Cartesian co-ordinate system, the latter of which we use in this work. In one dimension they are defined as:
\begin{equation}
    B_p(x; \beta) \equiv \beta^{-\frac{1}{2}} \left[2^p \pi^2 p\,! \right]^{-\frac{1}{2}} H_p(\beta^{-1}x)\exp\left(\dfrac{-x^2}{2\beta^{2}}\right),
\label{eq:basis_1D}
\end{equation}
where $\beta$ is a scaling factor, $p$ is a positive integer, and $H_p(x)$ is a hermite polynomial of order $p$. For our purposes, $x$ is the abscissa in a Cartesian co-ordinate system. These basis functions can be converted to two dimensions by multiplication:
\begin{equation}
    B_{p_1,p_2}(x,y ; \beta_1, \beta_2) = B_{p_1}(x; \beta_1)B_{p_2}(y; \beta_2),
\label{eq:basis_2D}
\end{equation}
where $y$ is the ordinate in a Cartesian co-ordinate system. The fourier transform $\Tilde{B}_{p_1,p_2}(k_x,k_y)$ of $B_{p_1,p_2}(x,y; \beta_1, \beta_2)$ is given by:
\begin{equation}
    \Tilde{B}_{p_1,p_2}(k_x,k_y) = i^{p_1 + p_2}B_{p_1,p_2}(k_x,k_y ; \beta_1^{-1}, \beta_2^{-1}),
\label{eq:basis_FT}
\end{equation}
meaning basis functions fitted in image space can be analytically calculated directly in $u,v$ space, as $k_x,k_y \propto u,v$. One further parameter of interest is the highest order basis function to fit for, $p_{\mathrm{max}} = p_1 + p_2$, which sets the maximum resolution that can be fitted for. Given a maximum (object size) and minimum angular extent (point spread function (PSF) resolution) to be fitted, $\vartheta_{\mathrm{max}},\vartheta_{\mathrm{min}}$, \citet{Refregier2003} show that good starting points for $\beta$ and $p_{\mathrm{max}}$ are
\begin{align}
   &\beta \approx (\vartheta_{\mathrm{min}}\vartheta_{\mathrm{max}})^{\frac{1}{2}} \quad \textrm{and} \\
   &p_{\mathrm{max}} \approx \dfrac{\vartheta_{\mathrm{max}}}{\vartheta_{\mathrm{min}}} - 1.
   \label{eq:basis_minmax}
\end{align}

The consequence of setting $p_{\mathrm{max}}$ is that the total number of basis functions to fit, $p_\mathrm{tot}$, grows as
\begin{equation}
p_\mathrm{tot} = \dfrac{(p_{\mathrm{max}}+2)(p_{\mathrm{max}}+1)}{2},
\end{equation}
meaning that the number of basis functions required grows quickly with increasing model resolution and size.

\subsection{Fitting with SHAMFI}
\label{subsec:fit_with_shamfi}
The simplest method to fit the basis functions in image space is to set up a number of linear equations, where for an image with $s \times t$ pixels, each pixel value $P$ at location $x,y$ can be described by:
\begin{equation}
    \begin{bmatrix} 
    B_{0,0}(x_0,y_0) & \sdots & B_{p_\mathrm{max},0}(x_0,y_0) \\
    \svdots &  \sddots & \svdots \\
    B_{0,0}(x_s,y_t) & \sdots & B_{p_\mathrm{max},0}(x_s,y_t)
    \end{bmatrix}
    \begin{bmatrix}
    C_{0,0} \\
    \svdots \\
    C_{p_\mathrm{max},0}
    \end{bmatrix}
    =
    \begin{bmatrix}
    P_{0,0} \\
    \svdots \\
    P_{s,t}
    \end{bmatrix}
\end{equation}
where $C_{p_1,p_2}$ is a scalar coefficient for the basis function $B_{p_1,p_2}$, for all basis functions where $p_1 + p_2 \leq p_{\mathrm{max}}$. This reduces the problem to the well-studied linear matrix $\mathbf{A}\mathbf{x} = \mathbf{b}$. For convenience, we use the \texttt{python} function \texttt{numpy.linalg.lstsq}\footnote{\url{https://docs.scipy.org/doc/numpy-1.13.0/reference/generated/numpy.linalg.lstsq.html}}, which solves for $\mathbf{x}$ by minimising the Euclidean 2-norm $|| \mathbf{b} - \mathbf{A} \mathbf{x} ||^2$.

We choose to fit the basis functions in image space, as it is far easier to isolate emission from a single object. Each visibility is the sum of the emission of all sources within the field of view, which is of particular concern for large field-of-view instruments such as the MWA. One can simply crop or edit in image space to remove difficult/unwanted pixels. \texttt{CLEAN}ed restored images offer an easy data set to fit, as the restoring beam tends to smooth out any sharp model pixelisation effects from the \texttt{CLEAN} fitting procedure. By design, these images are `true' sky emission convolved with a Gaussian kernel known as the `restoring beam', which is an approximation of the synthesised PSF. The synthesised PSF is the image-based manifestation of the incomplete sampling of an interferometer in the Fourier transform space in which it samples data, and is the reason images must be deconvolved. For a more accurate and instrument-independent shapelet fit, one must account for this. Rather than deconvolve the \texttt{CLEAN}ed image however, we achieve the equivalent by convolving the basis functions themselves by the restoring beam kernel, as outlined in the LOFAR cook-book mentioned in Section~\ref{sec:intro}.

Right-ascension ($RA$) and declination ($\delta$) are related to $x,y$ via a rotation by the position angle $\phi_\mathrm{PA}$, defined as the the increasing angle to East from North. Aside from fitting $C_{p_1,p_2}$,  $\beta_1$, $\beta_2$ and $\phi_\mathrm{PA}$ are all variables. Minimising $|| \mathbf{b} - \mathbf{A} \mathbf{x} ||^2$ for many values of $\beta_1,\beta_2$, and $\phi_\mathrm{PA}$ is computationally expensive. We instead settle for an initial 2D-Gaussian fit (defined in Equation~\ref{eq:gauss-def}) to the entire object to determine $\phi_\mathrm{PA}$, and then fit a range of $\beta_1,\beta_2$ in a grid, using the residuals of each fit to determine the optimal parameters. Examples of these steps are given in Sections~\ref{subsec:model_VLAsim} and~\ref{subsec:realmodel}.

\section{WODEN}
\label{sec:woden}
To test the \texttt{MS CLEAN} and shapelet fitting methods, we first use simulated data to isolate fitting errors from the plethora of calibration and instrumental effects present in real data. To enable a fair comparison of the speed of point source, Gaussian, and shapelet visibility model generation, we found it necessary to write a bespoke simulator (\texttt{WODEN}), which we detail here. This gave us control over the optimisation of each generation method, allowing consistency. We found when testing model generation within the \texttt{RTS} that the architecture of the code is optimised for shapelet models, rather than models built with multiple point/Gaussian components, due to the original needs of the calibration design at the inception of the \texttt{RTS}. Work is underway to update the architecture but remains on-going.

\subsection{Point source generation}
\label{subsec:wodenpointsource}

\texttt{WODEN} analytically generates a sky model for calibration directly in visibility space via the measurement equation~\citep[c.f.][]{Thompson2001}
\begin{multline}
V(u,v,w) =  \\ \int \mathcal{B}(l,m) I(l,m) \exp[-2\pi i(ul + vm + w(n-1))] \dfrac{dldm}{n},
\label{eq:measure_eq}
\end{multline}
where $V(u,v,w)$ is the measured visibility at baseline co-ordinates $u,v,w$, given the sky intensity $I(l,m)$ and instrument beam pattern $\mathcal{B}(l,m)$, which are functions of the direction cosines $l,m$, with $n=\sqrt{1-l^2-m^2}$.  Equation~\ref{eq:measure_eq} is discretised for point sources such that 
\begin{multline}
V(u_i,v_i,w_i) = \\ \sum_j \mathcal{B}(l_j,m_j)I(l_j,m_j) \exp[-2\pi i(u_il_j + v_im_j + w_i(n_j-1))],
\label{eq:measure_eq_discrete}
\end{multline}
where $u_i,v_i,w_i$ are the visibility co-ordinates of the $i^{\mathrm{th}}$ baseline, and $l_j$, $m_j$, $n_j$ is the sky position of the $j^{\mathrm{th}}$ point source.

This is an embarrassingly parallel problem, well suited to optimisation on GPUs. The basic application of a GPU to a problem such as this, is to calculate all iterations over $i,j$ in parallel, and then sum over $j$ after the fact. While this efficiently leverages the parallel computational structure of the GPU, it requires large arrays to store the $i \times j$ outputs, which can quickly use large amounts of GPU memory when $j$ becomes large. 

In recent years, the \texttt{CUDA} GPU-language~\citep{Nickolls2008} has made progress in making efficient memory managed read-modify-write operations on a single address, or so called \texttt{atomic} operations\footnote{\url{https://docs.nvidia.com/cuda/cuda-c-programming-guide/index.html\#atomic-functions}}. These allow calculations to be made in parallel, and the results written out to a single address with `hardware-level' thread-safe management, with minimal slow-down. We make use of \texttt{atomicAdd}, which adds the output of a thread to a single memory address, to perform the summation over $j$ in Equation~\ref{eq:measure_eq_discrete}. This reduces the size of the array needed to store outputs significantly.

We make no attempt to add any instrumental or ionospheric effects, and write \texttt{WODEN} to simply evaluate Equation~\ref{eq:measure_eq_discrete} with $\mathcal{B}(l,m) \equiv 1$ for a given array layout and observation parameters. We use \texttt{WODEN} here simply to test the modelling efficiency and model-generation costs of both the \texttt{MS} \texttt{CLEAN} and shapelet methods. 

\subsection{Gaussian and shapelet generation}
\label{subsec:wodengausspoint}

To simulate peeling using both shapelets and \texttt{MS CLEAN} outputs, we add the functionality to simulate Gaussians and shapelets into \texttt{WODEN}. We define a two-dimensional Gaussian in image space as:
\begin{equation}
G(x,y) = \exp \left( -4\ln(2)\left[\frac{(x - x_0)^2}{\theta_\mathrm{maj}^2} + \frac{(y - y_0)^2}{\theta_\mathrm{min}^2} \right] \right) 
\label{eq:gauss-def}
\end{equation}
where $x_0,y_0$ is the central point, and $\theta_\mathrm{maj},\theta_\mathrm{min}$ are the full-width half-maximum (FWHM) major and minor axes. Right-ascension and declination are related to $x,y$ via a rotation by the position angle $\phi_\mathrm{PA}$.
We utilise the \texttt{RTS} methodology of inserting a visibility `envelope' $\xi$ into Equation~\ref{eq:measure_eq_discrete} to create a Gaussian or shapelet source:
{\small
\begin{multline}
V(u_i,v_i,w_i) = \\
\sum_j \xi_j(u_i,v_i)I(l_j,m_j) \exp[-2\pi i(u_il_j + v_im_j + w_i(n_j-1))],
\label{eq:env_visi_def}
\end{multline}
}
where for a Gaussian:
\begin{align}
&\xi_j = \exp\left( -\dfrac{\pi^2}{4\ln(2)} \left( k_x^2\theta_{\mathrm{maj}j}^2 + k_y^2\theta_{\mathrm{min}j}^2\right) \right); \label{eq:gauss-env} \\
&k_x =  \cos(\phi_{PAj})v_i + \sin(\phi_{PAj})u_i; \label{eq:scale-gauss-x} \\
&k_y = -\sin(\phi_{PAj})v_i + \cos(\phi_{PAj})u_i; \label{eq:scale-gauss-y}
\end{align}
(noting that as visibilities are the Fourier dual of image space, the normalisation factor is inverted from Equation~\ref{eq:gauss-def}) and for a shapelet:
\begin{align}
&\xi_j = \sum^{p_k +p_l < p_\mathrm{max}}_{k,l} C_{p_k,p_l} \Tilde{B}_{p_k,p_l}(k_x,k_y); \label{eq:shape-env} \\
&k_x =  \dfrac{\pi}{\sqrt{2\ln(2)}} \left[\cos(\phi_{PA})v_{i,j} + \sin(\phi_{PA})u_{i,j} \right]; \label{eq:scale-shape-x} \\
&k_y = \dfrac{\pi}{\sqrt{2\ln(2)}} \left[-\sin(\phi_{PA})v_{i,j} + \cos(\phi_{PA})u_{i,j} \right], \label{eq:scale-shape-y}
\end{align}
where $u_{i,j},v_{i,j}$ are visibility co-ordinates for baseline $i$, calculated with a phase-centre $RA_j,\delta_j$, which corresponds to the central position $x_0,y_0$ used to fit the shapelet model in image-space. Inserting the visibility envelope $\xi$ in this way causes a convolution in visibility space, which in turn causes a multiplication in image space. Using Equation~\ref{eq:gauss-def} causes a convolution with a Gaussian kernel that sums to one, meaning the visibility can simply be multiplied by $I(l,m)$ to give the correct flux density. When using Equation~\ref{eq:shape-env}, one must scale the coefficients $C_{p_k,p_l}$ to ensure the kernel sums to one, to use Equation~\ref{eq:env_visi_def} for both Gaussian and shapelet components. In Equation~\ref{eq:shape-env} then we set $C_{p_k,p_l} \equiv C_{p_k,p_l} / I(l_j,m_j)$, and note that the coefficients output by \texttt{SHAMFI} have this division by integrated flux density applied, hence Equation~\ref{eq:shape-env} is implemented in \texttt{WODEN} exactly as it appears.

The shapelet basis function values $\Tilde{B}_{p_k,p_l}(u,v)$ can be calculated by interpolating from one dimensional look-up tables of $\Tilde{B}(k_x;1)$ (c.f. Equation~\ref{eq:basis_1D}), and scaling by the appropriate $\beta$. When using the scalings presented in Equations~\ref{eq:scale-gauss-x}, \ref{eq:scale-gauss-y}, \ref{eq:scale-shape-x}, and \ref{eq:scale-shape-y}, the first order shapelet basis function is equal to a 2D Gaussian when $\beta_1=\theta_\mathrm{maj}, \beta_2=\theta_\mathrm{min}$, explicitly $G(x,y,\theta_\mathrm{maj},\theta_\mathrm{min}) \equiv B_{0,0}(x,y,\theta_\mathrm{maj},\theta_\mathrm{min})$. This is useful to test the consistency between simulating Gaussians and shapelets.



For all simulation bench-marking used in this paper we use a GeForce GTX 1080 Ti NVIDIA GPU, with 12$\,$GB of memory. This allows us to store all time steps in GPU memory, meaning the optimisation over $j$ includes all time steps. The net result of these optimisations is that per frequency channel, we launch a single \texttt{CUDA} kernel instance each for all point sources, Gaussians, and shapelets, all of which are optimised with a consistent strategy. As $u,v,w$ only differ in frequency by a scaling factor, we only calculate them once per time step.

\section{Simulating and modelling Fornax~A}
\label{sec:sim_forA}

\subsection{VLA Fornax A Model}
\label{subsec:VLAmodel}
\begin{figure}[htb]
\centering
\includegraphics[width=0.85\columnwidth]{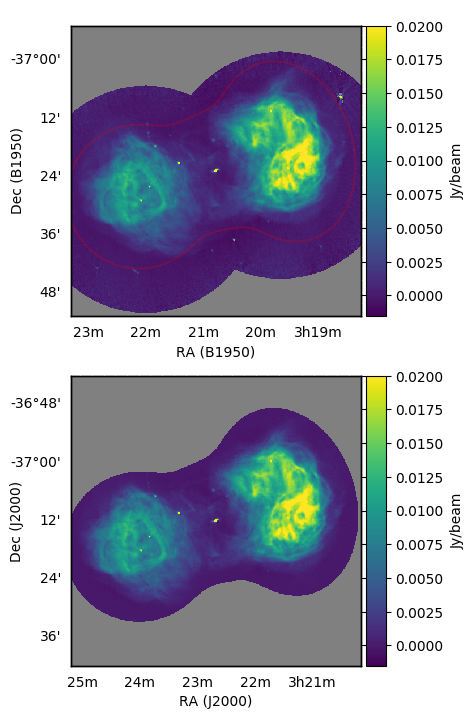}
\caption{\textit{Upper panel:} Fornax~A image obtained from NED, which includes \texttt{CLEAN} residuals around the edge of the lobes. Over-plotted in red is a boundary line outside of which a Gaussian taper was applied to create the model used in WODEN. \textit{Lower panel:} The image after applying the taper and transforming from B1950 to J2000 centred co-ordinates. Any pixels that were not masked (the grey region) in the lower image were converted into point sources.}
\label{fig:VLA_forA}
\end{figure}

The goal of this work is to compare shapelet and \texttt{MS CLEAN} methods to generate models of extended radio galaxies. To fairly test the two methods as implemented in \texttt{SHAMFI} and \texttt{WSClean}, in reality one would start with calibrated visibility data, deconvolve and image using \texttt{WSClean} to create \texttt{MS CLEAN} components, and create an image to use when modelling with \texttt{SHAMFI}. This would ensure both methods use the same data set, with the same angular resolution, and both suffer any systematics inherent to the data. As stated in Sections~\ref{sec:intro} and~\ref{sec:woden}, we choose to test the methods first using simulated data, to reduce the number of systematics. To do this, we need to create simulated `observed' data in visibility space.
 We choose to simulate observed visibility data using point sources alone, to minimise biasing towards either method. To build a realistic and high-resolution model of a complex and extended radio galaxy, we use a public VLA image of Fornax~A at 1.4~GHz, obtained from the NASA/IPAC Extragalactic Database\footnote{\url{http://ned.ipac.caltech.edu/uri/NED::Image/fits/1989ApJ...346L..17F/NGC_1316:I:20cm:fev1989:i}}, a digitised version of the data presented in~\citet{Fomalont1989}. The online FITS file is a cutout around the galaxy, containing some \texttt{CLEAN} residuals. To create a point source model, we first apply a Gaussian taper around the edge of the image (see Figure~\ref{fig:VLA_forA}). Not only does this reduce \texttt{CLEAN} residuals around the image, but also removes the sharp boundary edge, minimising future ringing when imaging from the simulated visibility data. We then move the image from B1950 to J2000 co-ordinates, convert each pixel into a point source (explicitly we convert each pixel into a Dirac delta function, allowing us to run Equation~\ref{eq:measure_eq_discrete} over all pixels), and scale the flux density of all pixels to sum 500$\,$Jy at 180$\,$MHz. We give each pixel a single spectral index of $-0.8$. As the restoring beam of this \texttt{CLEAN}-ed image is smaller than the resolution of the instruments used in simulation, we make no attempt to remove it from the image.

\begin{figure}[htb]
\centering
\includegraphics[width=0.85\columnwidth]{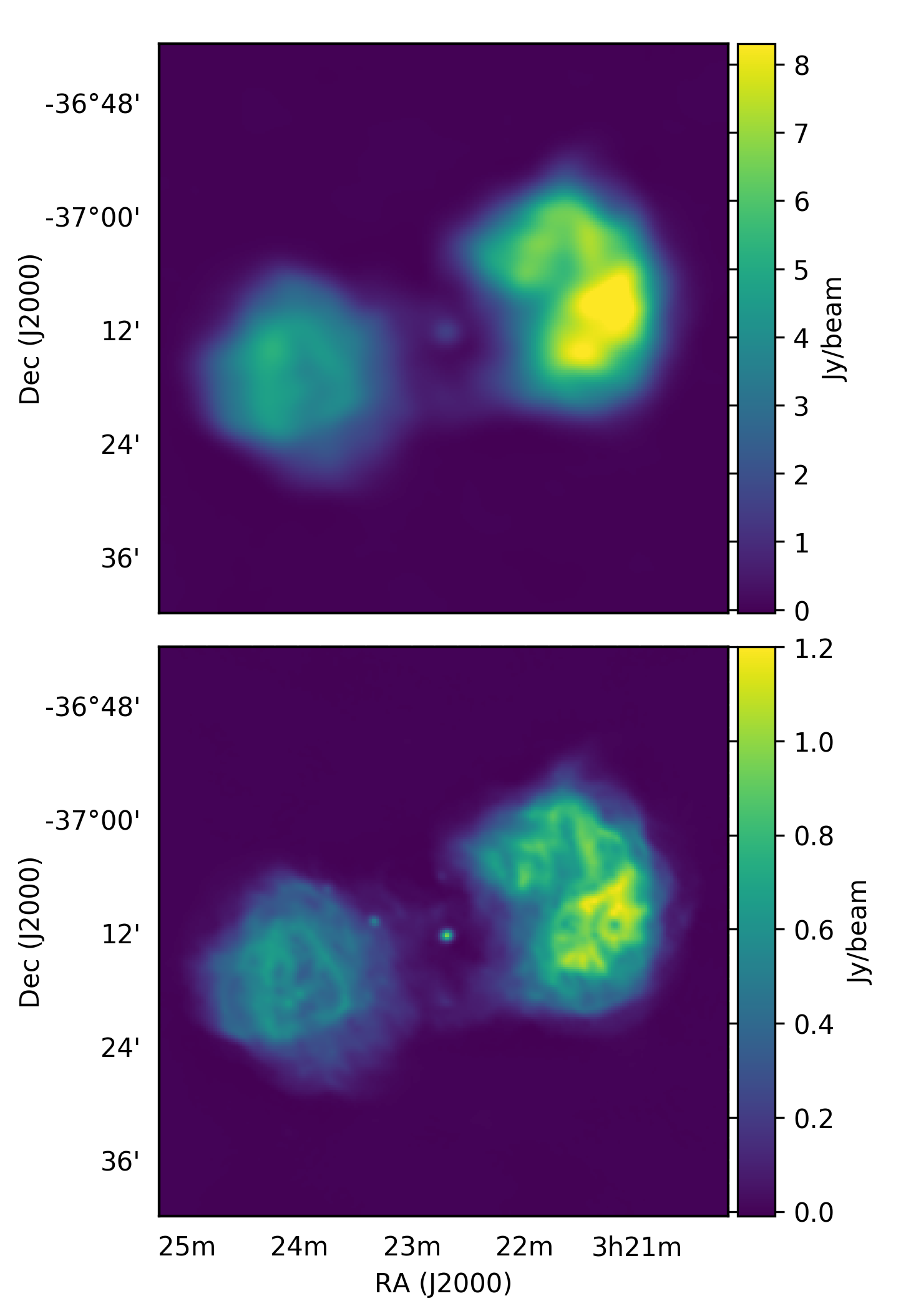}
\caption{\texttt{MS CLEAN}ed images of the \texttt{WODEN} simulation of Fornax~A. \textit{Top}: All phase I configuration, imaged with briggs 0 weighting to balance the large number of short baselines present in phase I data with a reasonable resolution. \textit{Bottom}: A combination of both phase I and phase II configurations. These are imaged with uniform weighting to take advantage of the higher resolution phase II data.
\label{fig:sim_nocal_phase1-phase2}
}
\end{figure}

We simulate 6 MWA observations of 30.72$\,$MHz bandwidth and 10$\,$kHz spectral resolution using \texttt{WODEN}. These simulations use the observational parameters of the real data used to image Fornax~A in Section~\ref{subsec:realmodel}. As the MWA has recently been upgraded with added receiver elements, dubbed phase II, it can be operated in various configurations, with different angular resolutions and sensitivities to diffuse or compact emission~\citep[see][for details]{Wayth2018}. Three of the observations use the phase I configuration (maximum baseline $\sim3\,$km), with the other three the phase II extended configuration (maximum baseline $\sim5.5\,$km, hereto referred to as phase II). The phase I observations have 2$\,$s time resolution and the phase II observations 0.5$\,$s resolution to reduce time decorrelation on the longer baselines. We simulate 56 time steps for all observations regardless of the time resolution to reduce computational load. We use \texttt{WSClean} to perform a \texttt{MS CLEAN} on the simulated data, with the results shown in Figure~\ref{fig:sim_nocal_phase1-phase2}. These images have not been beam corrected, given that \texttt{WODEN} does not include beam effects. We include an example \texttt{WSClean} command in Figure~\ref{code:wsclean_command} to detail the settings used.

\subsection{Modelling the Fornax A simulation}
\label{subsec:model_VLAsim}

We fit the simulated Fornax A in the images in Figure~\ref{fig:sim_nocal_phase1-phase2} using \texttt{SHAMFI} as described in Section~\ref{sec:shape_fitting}, and use Equation~\ref{eq:basis_minmax} to set $p_{\mathrm{max}}$. We set $\vartheta_{\mathrm{max}} = 0.5^{\circ}$, and set $\vartheta_{\mathrm{min}}$ by oversampling the angular resolution of the array by 3, giving $\vartheta_{\mathrm{min}} = \lambda / 3b_{\mathrm{max}}$, where $\lambda$ is the wavelength of the observation, and $b_{\mathrm{max}}$ the maximum baseline of the array. We set the latter to 3$\,$km for the phase I array and 5.5$\,$km for the phase II array. This yields $p_{\mathrm{max}} = 46$ for phase I and $p_{\mathrm{max}} = 86$ for phase II.

\begin{figure}[htb]
\centering
\includegraphics[width=1.0\columnwidth]{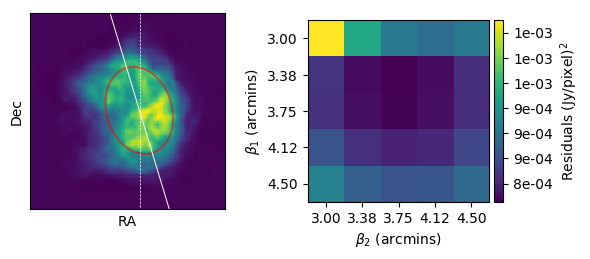}
\caption{\textit{Left:} Example of initial Gaussian fit used to set $\phi_{\mathrm{PA}}$ of the basis functions for one of the simulated phase 1+2 lobe images. The red line shows the FWHM, with the white lines demonstrating the PA found. \textit{Right:} Example of the grid based approach for fitting $\beta_1$ and $\beta_2$. The colour scale here represents the residuals in (Jy/pixel)$^2$ left after fitting the image in the left panel with all basis functions up to $p_{\mathrm{max}} = 86$.}
\label{fig:fit_PA}
\end{figure}

\begin{figure*}[htb!]
\centering
\includegraphics[width=0.9\textwidth]{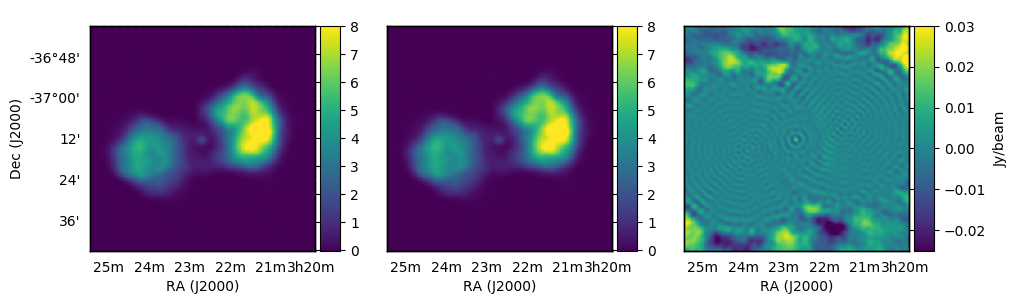}
\caption{\textit{Left:} \texttt{MS CLEAN}ed phase 1 simulated data \textit{Middle:} Fitted shapelet model recreated in image space, using restoring beam convolved basis functions \textit{Right:} The model subtracted from the data. The footprints of the two separate shapelet models (one for each lobe) are clearly visible in the residual plot.}
\label{fig:sim-model_phase1}
\end{figure*}

During modelling we found the choice of location of the zero pixel $x,y = 0,0$ of the basis functions severely affected the quality of fit. We found that for a very extended, double-lobed radio galaxy like Fornax~A, the best solution was to split the galaxy into two lobes, and fit each lobe separately. To avoid double-fitting flux, we divided the image in two by fitting each lobe with a normalised two-dimensional Gaussian, which we label $N_1(x,y)$, $N_2(x,y)$, and then applying weights $w_1(x,y),w_2(x,y)$ to each pixel such that $w_1 = N_1 / (N_1 + N_2)$ , $w_2 = N_2 / (N_1 + N_2)$. Figure~\ref{fig:fit_PA} shows an example of fitting for $\phi$ and the grid-search optimising for $\beta_1,\beta_2$. The fitting results for the simulated phase I data are shown in Figure~\ref{fig:sim-model_phase1}.

As is also prescribed in the LOFAR cookbook, we found it necessary to subtract any compact point source-like emission from the images including phase II data, which reduced the number of higher order shapelets required for a good fit. We do this manually by subtracting image-based Gaussian components, as we found traditional source finders struggle to deal with the complexity of Fornax~A to successfully isolate the point-like details in the image. A full example of the image preparation needed to model Fornax A is given in Section~\ref{subsec:realmodel}. We note the tools to split a radio galaxy into Gaussian masked regions, and manually subtract Gaussians, are included in the \texttt{SHAMFI} package.

\section{Simulated peeling results}
\label{sec:sim_peel-results}
The main motivation behind this paper is to understand how modelling Fornax~A might impact a PS estimation of the EoR through residuals left after peeling. We therefore only test peeling on phase I data here, as the phase II layout does not have the necessary short baselines required to detect the EoR signal. To mimic peeling, we simulate the full point source model from Section~\ref{subsec:VLAmodel}, the \texttt{MS CLEAN} component models from the images in Figure~\ref{fig:sim_nocal_phase1-phase2}, and the fitted shapelet models described in Section~\ref{subsec:model_VLAsim}, with the same time and frequency cadence ($8\,$s, 80$\,$kHz), all phase-centred on Fornax~A, for a single 2 minute observation. We then fit a complex gain per tile (as we simulate no beam, we have no polarisation, and so only need to fit a single complex gain) for each set of \texttt{MS CLEAN} and shapelet visibilities to the full simulation, apply the gains to the model, and subtract from the full simulated visibilities. While no antenna gains were ever added during the simulation, real peeling includes a calibration step, so we include that step in our analysis. We reiterate that we have also not added thermal noise to the simulations, making this a strongly idealised case. 

\subsection{Peeling method on simulated data}
\label{subsec:sim_peel_method}
To fit the gains, we follow the calibration scheme implemented in \texttt{YANDAsoft}\footnote{\url{https://github.com/ATNF/yandasoft}}. As this scheme has yet to be published we detail the formalism here. We label the ideal visibility model for baseline $ij$ as $\mathcal{I}_{ij}$. The visibility model $M_{ij}$ (over the range of time and frequency that antenna gains are assumed constant) is
\begin{equation}
    M_{ij}(t,\nu) = g_ig_j^* \mathcal{I}_{ij}(t,\nu).
\end{equation}
where $g_i$ is the gain on antenna $i$. The measured visibility $V_{ij}$ is
\begin{equation}
    V_{ij}(t,\nu) = (g_i+e_i)(g_j+e_j)^*\mathcal{I}_{ij}(t,\nu),
\end{equation}
where $e_i$ is an additive error on the gain $g_i$. We label the residual visibilities as
\begin{equation}
    R_{ij}(t,\nu) = V_{ij}(t,\nu) - M_{ij}(t,\nu).
\end{equation}
The gains can be optimised by iteratively solving for and following local error gradients. These are found using linear least-squares with data $\Re(R)$ and $\Im(R)$ and free parameters $\Re(e)$ and $\Im(e)$. The complex conjugation of gains in the equations above cannot be simply represented by a linear operator. So rather than solving for complex solutions, we instead solve for the real and imaginary parts separately and multiply the relevant coefficients by $-1$. Assuming the error-squared term is small allows the problem to be linearised as $\mathbf{A}\mathbf{x} = \mathbf{b}$, where $\mathbf{A}$ is the design matrix, $\mathbf{x}$ are the parameters, and $\mathbf{b}$ is the residual vector, e.g the real parts of the residual vector $\mathbf{b}$ are
\begin{equation}
    \Re[R_{ij}(t,\nu)] = \Re[ e_j^*g_iI_{ij}(t,\nu) + e_ig_j^*I_{ij}(t,\nu)].
\end{equation}
This can be implemented to solve for all $e_i$ in the vector $\mathbf{x}$ as
\begin{equation}
\mathbf{x} = \left[ \mathbf{A}^\intercal \mathbf{A} \right]^{-1}\mathbf{A}^\intercal \mathbf{b}.
\label{eq:peeling-iter}
\end{equation}
The assumption of linearity will improve as the system converges. We found that as the simulations had no instrumental or source sidelobe noise, the gains consistently converged after 10 iterations of the calibration loop, and so used 10 iterations for all the results shown in this and following Sections. We fit a separate gain for each time and frequency step. Once we have the gains, we then fit a cubic spline across the entire bandwidth using the \texttt{scipy.interpolate.interp1d}\footnote{\url{https://docs.scipy.org/doc/scipy/reference/generated/scipy.interpolate.interp1d.html}} function, to ensure we apply spectrally-smooth gains during peeling. This was found to be necessary for an EoR detection by~\citet{MouriSardarabadi2019}, as without smooth gains, the EoR signal can be suppressed. The caveat to this being that we are assuming a linearised calibration case, where spectrally-smoothing the gains does yield improvement. In the non-linear regime this may not be true, however given that these simulations have zero noise and instrumental effects, the error terms in calibration are likely small.

\subsection{Peeling results}
\label{subsec:sim_peeling_results}

\begin{figure*}[htb]
\centering
\includegraphics[width=0.90\textwidth]{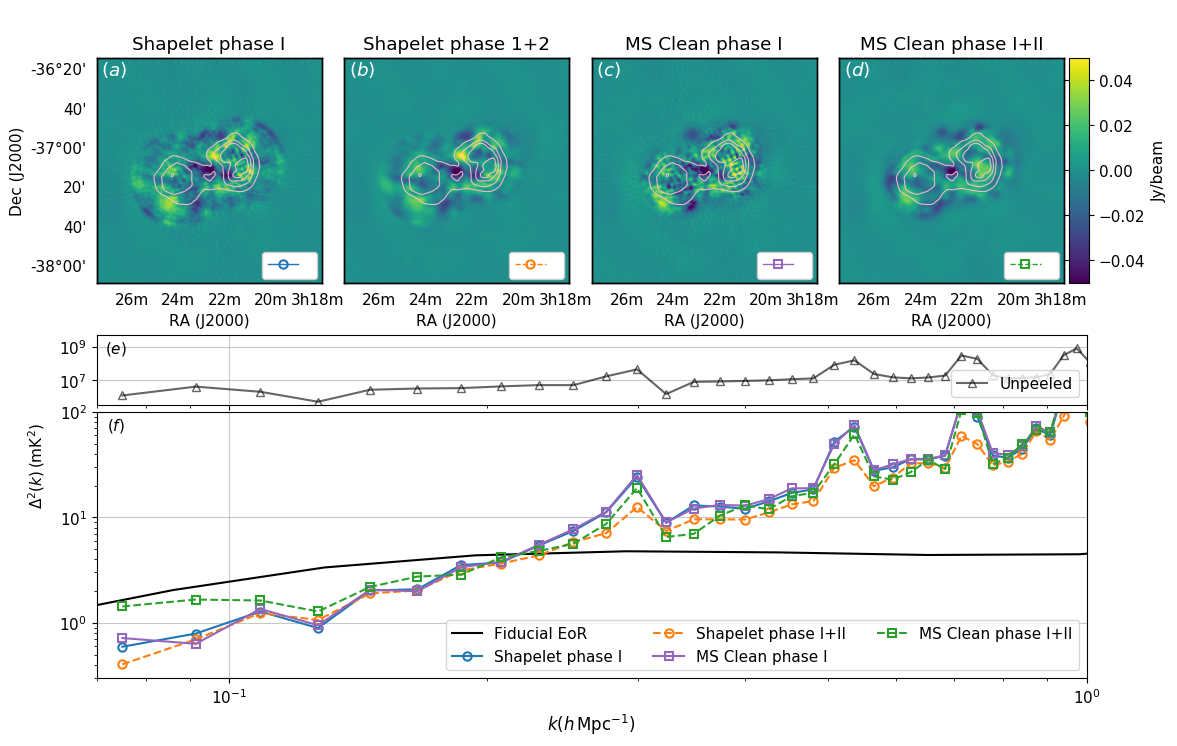}
\caption{Simulated peeling results. Plots ($a$) to ($d$) show \texttt{MS CLEAN}ed peel residuals of an integration over the full bandwidth and time span of a single simulated phase 1 observation, each peeled with a different model for Fornax~A. The models are: ($a$) a shapelet model from phase I data; ($b$) a shapelet model from phase 1 and 2 data; ($c$) a \texttt{MS CLEAN} model from phase I data; ($d$) a \texttt{MS CLEAN} model from phase I and II data. A contour plot of the unpeeled power is shown on each plot as a angular scale reference. Plot ($f$) shows the 1D PS as estimated with CHIPS for the models in ($a$) to ($d$). For reference, the power without peeling is shown for the unpeeled \texttt{WODEN} simulation in ($e$). The line style and marker for each 1D PS plotted in ($f$) is also plotted in each plot from ($a$) to ($d$) on the lower right for reference.}
\label{fig:peel_phase1-vs-2}
\end{figure*}

\begin{figure*}[htb]
\centering
\includegraphics[width=0.85\textwidth]{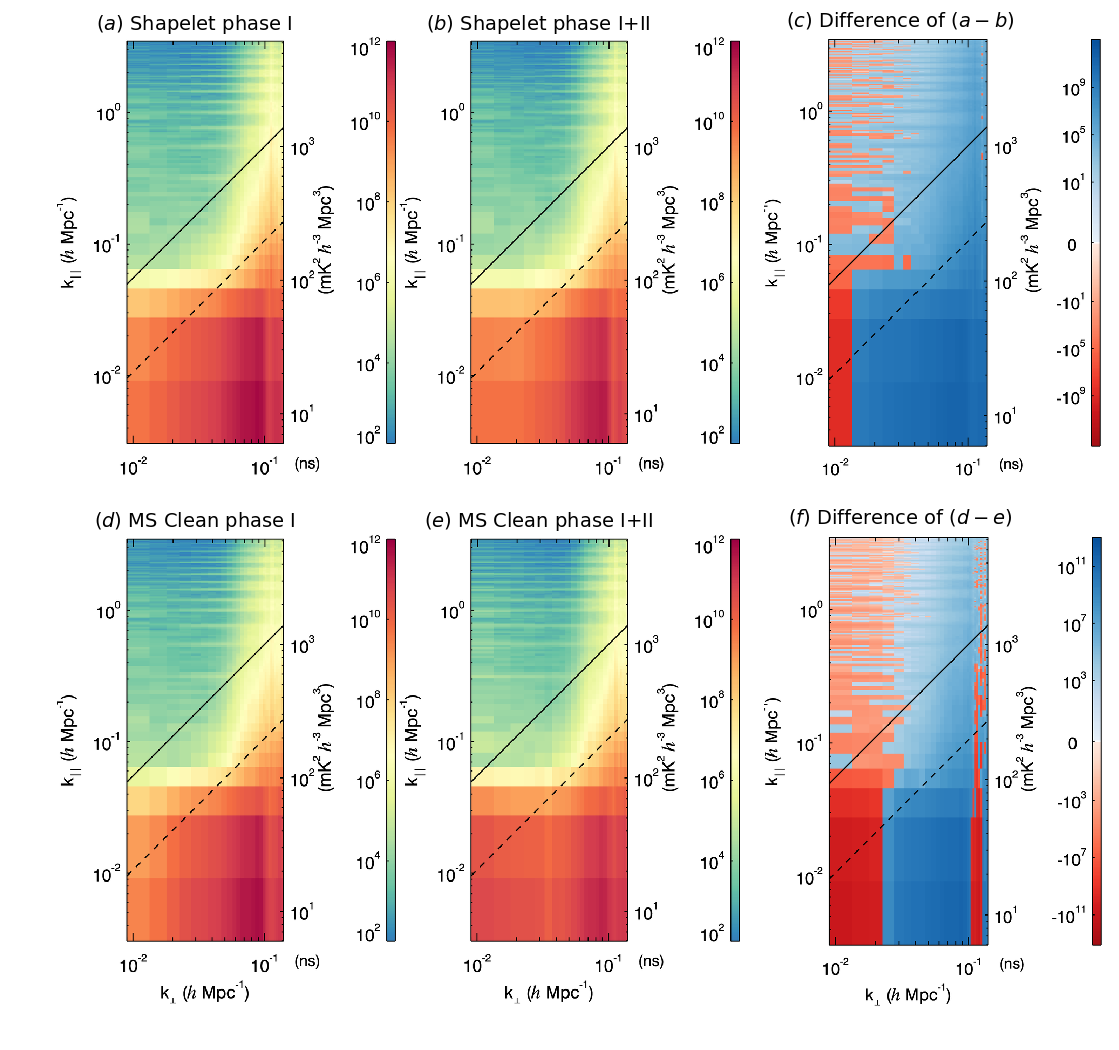}
\caption{2D PS as estimated by CHIPS of the simulated peeling results. Plots $(a)$ and $(b)$ show the shapelet model peel results for phase I and phase I+II, with $(c)$ showing the difference between the 2D PS in $(a)$ and $(b)$. The \texttt{MS CLEAN} peel results are shown in a similar manner in plots $(d)$, $(e)$, and $(f)$. In the difference PS of $(c)$ and $(f)$, blue means the phase I model subtracted less power than the phase I+II model, and red means the phase I model subtracted more power.}
\label{fig:peel_phase1-vs-2_2D}
\end{figure*}

The results of peeling all four models from the full simulation are shown in Figures~\ref{fig:peel_phase1-vs-2} and~\ref{fig:peel_phase1-vs-2_2D}. We image the peeling residuals in Figure~\ref{fig:peel_phase1-vs-2} using the same \texttt{WSClean} commands used to image the full simulation in Figure~\ref{fig:sim_nocal_phase1-phase2}. We make one-dimensional (1D, Figure~\ref{fig:peel_phase1-vs-2}) and two-dimensional (2D, Figure~\ref{fig:peel_phase1-vs-2_2D}) PS estimates using \texttt{CHIPS}~\citep{Trott2016}, and note that the values of the powers in Figures~\ref{fig:peel_phase1-vs-2} and~\ref{fig:peel_phase1-vs-2_2D} are indicative, rather than absolute. This is because \texttt{CHIPS} grids using the primary beam shape of the MWA, and normalises the outputs based on the expected cosmological observing volume of the telescope. Neither are present in \texttt{WODEN} simulations. All PS estimates were run with identical settings however, and so all differences are solely due to the models used for peeling.

The 1D PS is an average over the fourier inverse (or $k$-space) of three dimensions; two angular dimensions across the sky, and one derived from the spectral dimension of the data. The 1D PS is currently used to generate upper limits on an EoR detection. In Figure~\ref{fig:peel_phase1-vs-2} we include a fiducial EoR signal as a point of reference to the reader. This is taken from the public website\footnote{\url{http://homepage.sns.it/mesinger/EOS.html}} associated with the work of~\citet{Mesinger2016}. We select a redshift that matches the centre of the simulated frequency bandwidth. The greatest ratio of expected EoR signal to astrophysical foregrounds, after foreground removal/avoidance, is expected to lie at low $k$-modes in the 1D PS. In Figure~\ref{fig:peel_phase1-vs-2} we concentrate on the smallest $k$-modes. We find little difference when peeling with the phase I \texttt{MS CLEAN} and shapelet models (purple line with squares, and blue line with circles, respectively). We see a difference in behaviour in the phase I+II models however; including phase II data in the model improves the subtraction at small $k$ with the shapelet model, but actually worsens the subtraction with the \texttt{MS CLEAN} model.

To understand the difference, we turn to the 2D PS, shown in Figure~\ref{fig:peel_phase1-vs-2_2D}. The 2D PS averages the $k$-modes obtained from angular scales upon the sky ($k_\perp$, horizontal axis), and plots them against the $k$-modes derived from the spectral response ($k_\parallel$, vertical axis). Broadly, astrophysical foregrounds are expected to be spectrally smooth, and so should inhabit the bottom of the 2D PS, where $k_\parallel$ is small. Instrumental chromaticity causes the foregrounds to be swept up at higher $k_\perp$ into the so-called `wedge' on the lower right, with the upper left being known as the `window', offering the cleanest area of $k$-space in which to attempt a detection. If the peeled residuals were perfectly spectrally smooth, we would expect this window to be devoid of emission, however there is clearly power present. There are two processes that could have added in spectral structure: the calibration step during peeling, and the gridding kernel of \texttt{CHIPS}, which expects the data to have the imprint of an MWA beam pattern, which was not included in the simulation. Visibilities with quickly-changing spectral structure have the effect of throwing power that would normally sit at low $k_\parallel$ up into high $k_\parallel$, contaminating the window. This effect can happen in real data, and the net effect is more power in the wedge at a certain $k_\perp$ will also throw more power into the window at that $k_\perp$.

This effect goes some way to explaining the difference in 1D power when adding higher angular resolution data into the modelling process between shapelets and \texttt{MS CLEAN}. When looking at the difference of the phase I and phase I+II shapelet model residuals ($(c)$ Figure~\ref{fig:peel_phase1-vs-2_2D}), we see more power has been subtracted out of the wedge at all $k_\perp$ beside the very smallest $k_\perp$-bin, which has meant there is less power in the window also. Comparing the \texttt{MS CLEAN} models, the introduction of higher-angular resolution data has yielded a large improvement in the power subtracted at large $k_\perp$ (small angular scales), but has worsened the subtraction at low $k_\perp$, contaminating the window. It is difficult to isolate what might cause worsening subtraction at large angular scales with \texttt{MS CLEAN}. It is possible that as including the phase II configuration reduces the angular size and shape of the synthesised point spread function, an \texttt{MS CLEAN} is more biased to smaller angular scales. There are two \texttt{WSClean} parameters that can be manually set to change the \texttt{MS CLEAN settings}: \texttt{-multiscale-scales}, which forces the \texttt{CLEAN} to specific scales, and \texttt{-multiscale-scale-bias}, which biases the \texttt{CLEAN} to larger or smaller scales. We tested forcing the scales that were \texttt{CLEAN}-ed for the phase I+II image to be the same as the scales \texttt{CLEAN}-ed for the phase I alone (plus an extra smaller scale to account for the increase in resolution). We also changed \texttt{-multiscale-scale-bias} to favour larger scales. We propagated both these changes through to the PS, and found either no change in residuals, or worsened residuals at all scales. While not an exhaustive parameter search, we find the default settings give the best results, and so use them throughout the paper. What is clear is that the \texttt{SHAMFI} model was able to incorporate the increased angular resolution, while simultaneously still fitting the large scale structure.

\subsection{Model truncation}
\label{subsec:compress}
As discussed in Section~\ref{sec:sim_peel-results}, calculating a value for a single shapelet basis function requires more calculation than for a single point source or Gaussian. As a point of reference, for the simulated phase I + II image, \texttt{WSClean} produced a total of 6331 \texttt{MS CLEAN} components (4540 point source and 1791 Gaussian). Fitting a separate shapelet model for each lobe, each with $p_\mathrm{max} = 86$, gives a total of 7656 basis functions to extrapolate. Therefore, for shapelets to be computationally-competitive, some compression/truncation of the full model is necessary.
We apply a basic truncation, by simply removing the \texttt{MS CLEAN} components or shapelet basis functions that contribute the smallest absolute flux density to the total number of components $N_\mathrm{tot}$, up until some fraction $f_\textrm{comp}$ of all components remain. Explicitly, we find the number of components $N_\textrm{comp}$ that satisfies
\begin{align}
&\sum^{N_\textrm{comp}}_j |I_j| <= f_\textrm{comp} \sum^{N_\textrm{tot}}_j |I_j|, \\
\textrm{where}\quad\quad & |I_j| > |I_{j+1}|.
\end{align}
For a point source or Gaussian, $I_j$ is simply the total integrated flux density, with the absolute value necessary as \texttt{MS CLEAN} components and shapelets can be positive or negative. To find $I_j \equiv I_{p_1,p_2}$ for the shapelet basis function $B_{p_1,p_2}$, we image and sum the absolute of the restoring beam $G(\theta_{\mathrm{bmaj}},\theta_{\mathrm{bmin}})$ convolved basis function,
\begin{equation}
    I_{p_1,p_2} =  \sum_s \sum_t \left| C_{p_1,p_2} G(\theta_{\mathrm{bmaj}},\theta_{\mathrm{bmin}}) \ast B_{p_1,p_2}(x_s,y_t) \right|,
\end{equation}
where $\ast$ is a convolution, for all $s \times t$ pixels used during the model fit.

We have implemented this type of truncation within \texttt{SHAMFI}, which has the necessary code base to allow the compressed (reduced) set of shapelet basis functions to be fit to the image again. We apply this extra step in the following results.

\begin{figure*}[htb]
\centering
\includegraphics[width=0.99\textwidth]{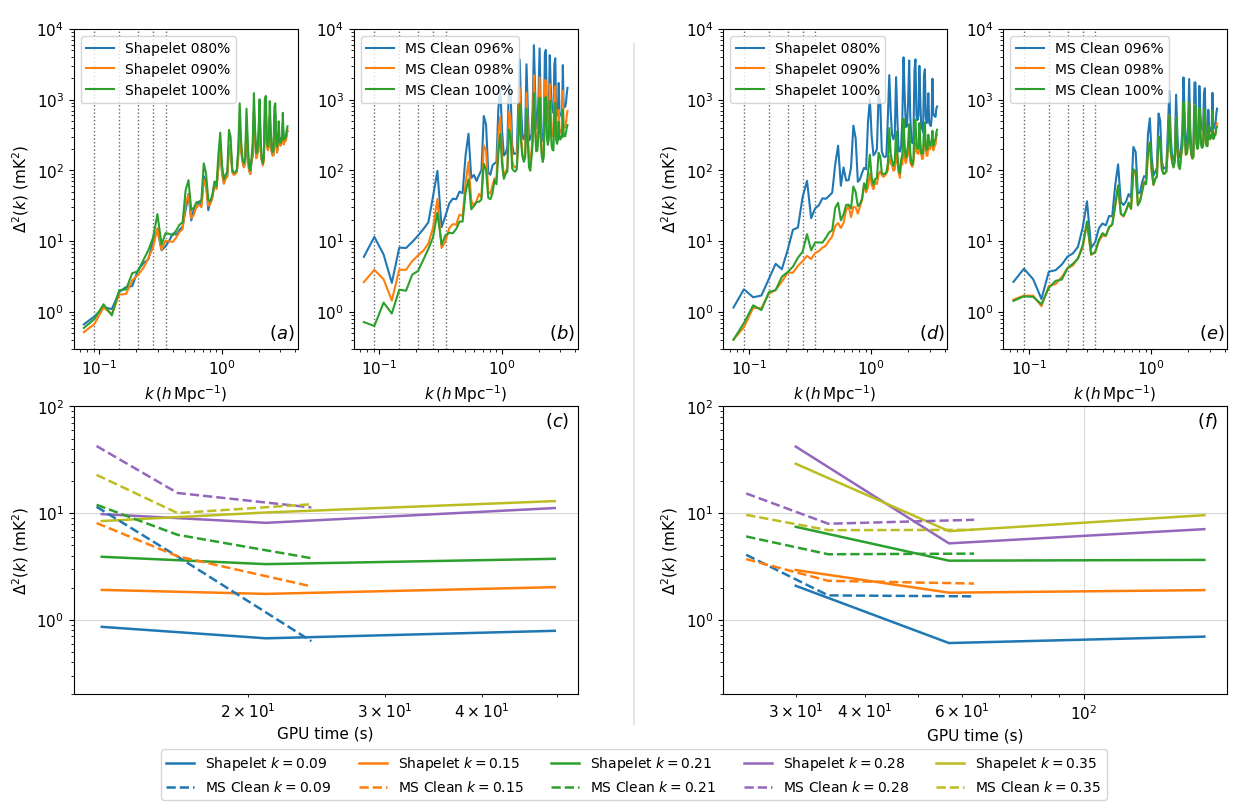}
\caption{Results from compressing the \texttt{MS CLEAN} and shapelet models. Plots $(a)$, $(b)$, $(c)$ on the left show results from phase I models, where $(d)$, $(e)$, $(f)$ on the right show results from phase I+II models. The top row show residual 1D PS left after peeling \texttt{MS CLEAN} models ($a$ and $d$) and shapelet models ($b$ and $e$), at various levels of truncation. The percentage in the legend is the percentage of remaining components after truncation. We take slices through the truncation results at specific low $k$ values, shown by the vertical dotted lines, and plot them as a function of GPU time in the bottom row ($c$ and $f$). In both plots, dashed lines are \texttt{MS CLEAN}, and solid lines are shapelet results. Matching colours plot matching $k$-modes.}
\label{fig:compress_results}
\end{figure*}

We apply three levels of truncation to both \texttt{MS CLEAN} and shapelets, for both the phase I and phase II models. We then simulate these models through \texttt{WODEN} for the same 2 minute observation as used in Section~\ref{sec:sim_peel-results}, and apply the same peeling technique, fitting a single complex gain per antenna with 10 iterations of Equation~\ref{eq:peeling-iter}. When running the simulations, we used \texttt{nvprof}\footnote{\url{https://docs.nvidia.com/cuda/profiler-users-guide/index.html}}, the \texttt{NVIDIA} line profiler, to measure the time spent performing calculations and memory allocations on the GPU. For each level of truncation, we ran the peeled outputs through \texttt{CHIPS}, so we could compare the effect of peeling against time taken in generating the visibilities. The results are shown in Figure~\ref{fig:compress_results}.
As expected, the uncompressed shapelet model takes more than twice the time to generate than the \texttt{MS CLEAN} model. However, for both the phase I and phase I+II models, shapelets can be compressed to a level where they incur comparative computational expense, and still peel out more power at the largest spatial scales. By comparing the blue dashed lines in the bottom plots of Figure~\ref{fig:compress_results}, we can see that the 100\% \texttt{MS CLEAN} model subtracts more large-scale power when only including phase I information, (bottom left), rather than including both phase I+II. The amount of power subtracted by the shapelet model is comparable when incorporating just phase I information or both phase I+II.

\section{Real data results}
\label{sec:real_data}

\subsection{Imaging and modelling real MWA data}
\label{subsec:realmodel}
\begin{figure*}[h]
\centering
\includegraphics[width=0.9\textwidth]{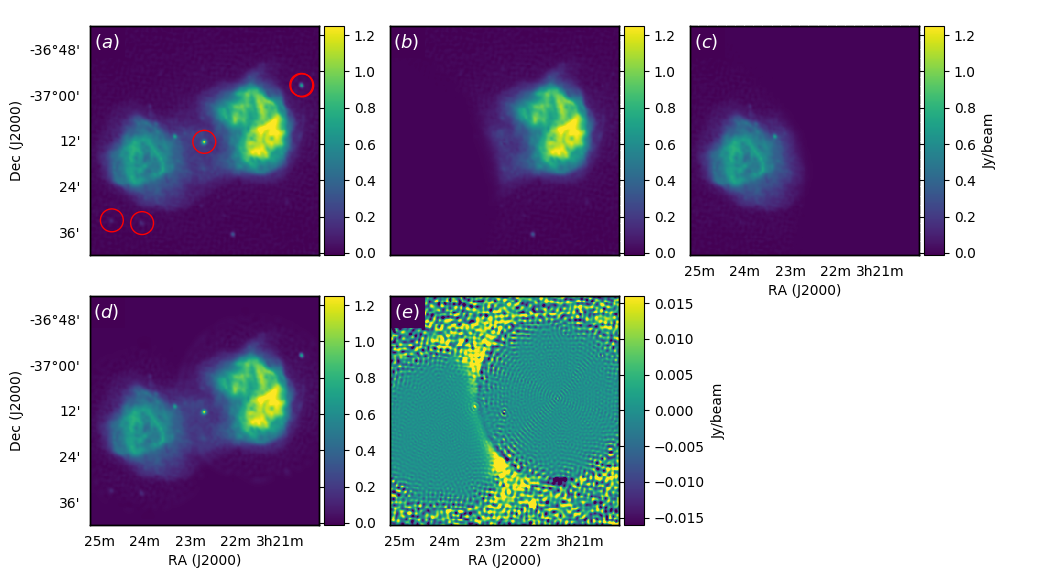}
\caption{Stages of fitting for the Fornax A image using real MWA data as described in Section~\ref{subsec:realmodel} and shown in $(a)$. Four compact sources (outlined in red circles) were found through trial and error to cause fitting problems, and so were subtracted as Gaussians. The image was then split into two lobes, $(b)$ and $(c)$, and fitted separately. $(d)$ shows the fitted shapelet model recreated in image space, using restoring beam convolved basis functions, as well as the subtracted Gaussians. $(e)$ shows the fitting residuals.}
\label{fig:fullmodel_real}
\end{figure*}

We use six 2$\,$minute MWA observations to create the image of Fornax A in Figure~\ref{fig:fullmodel_real}. Three observations were phase I, taken in December 2014 as part of the GLEAM survey~\citep{Wayth2015,Hurley-Walker2017}, and the other three were phase II, taken in February 2018 as part of the GLEAM-X survey~\citep[][in prep.]{Hurley-Walker2019}. Calibration observations of the bright radio galaxy Pictor~A were also taken at the same frequency on the same nights as the Fornax~A observations. The Pictor~A observations were used to solve for an initial set of calibration solutions, which were then applied to the Fornax~A data. An image was  made with the full 30.72~MHz bandwidth of the MWA, centered at 185$\,$MHz using joint deconvolution with \texttt{WSClean}. This joint deconvolution was enabled by the implementation of Image Domain Gridding~\citep[IDG,][]{VanderTol2018} into \texttt{WSClean}, which uses the MWA primary beam~\citep{Sokolowski2017} to properly take into account the changing beam shapes for each snapshot observation in the gridding process. Several rounds of self-calibration were then performed with \texttt{CALIBRATE}~\citep{Offringa2016} to improve the antenna gain solutions, using the \texttt{MS CLEAN} component model produced by \texttt{WSClean}. 

In Figure~\ref{fig:fullmodel_real} we detail the image manipulation required when running \texttt{SHAMFI} to obtain a good model using real data, including the necessary subtraction of compact emission to reduce the need for higher-order (and therefore higher-resolution) shapelet basis functions. The process of fitting the real Fornax A image, along with all commands used, are available online as a tutorial in the documentation of SHAMFI\footnote{\url{https://shamfi.readthedocs.io/en/latest/tutorial.html}}.

\subsection{Calibrating and peeling real data}
\label{subsec:calpeel_realdata}
To carry out calibration and peeling on the real data sets we use the \texttt{RTS}, software which is used by the Australian MWA EoR collaboration~\citep[see][for an overview of \texttt{RTS} processing]{Jacobs2016}. The \texttt{RTS} has the capacity to perform direction-dependent calibration, correct for first-order ionospheric effects, and peel sources~\citep[phase-rotate, calibrate, and subtract a source, see][]{Noordam2004}.
During the imaging of the real data, \texttt{WSClean} produced 2537 \texttt{MS CLEAN} components. These components, as well as the shapelet model described in Figure~\ref{fig:fullmodel_real}, were used to peel with the \texttt{RTS}. We peel Fornax~A from two data sets: the first set where Fornax~A is near the primary beam centre, and therefore contributes a large fraction of the total power in the sky; the second set made of standard MWA EoR observations, where Fornax~A is far from pointing centre, often at a total Stokes I beam power of $\sim 30\%$. The latter data set is pointed toward the `EoR1' field at $RA=4^h,\delta=-27^\circ$. We use the ionospheric metrics detailed in~\citet{Jordan2017} to pick an ionospherically-quiet night for the EoR1 data. The key points of both data sets are summarised in Table~\ref{table:obs_params}.
\begin{table}[h]
\centering
\caption{Observational parameters of the real data used to test peeling.}
\begin{tabular}{c | c | c}
\hline
Observation Centre & Fornax~A & EoR1 \\
\hline
\hline
Dates & 4$^{\mathrm{th}}$/12/2015 & \\
 & 6$^{\mathrm{th}}$/12/2015 & 29$^{\mathrm{th}}$/11/2015 \\
Central $\nu$ (MHz) & 154.255 & 182.415 \\
Time resolution (s) & 0.5 & 2 \\
$\nu$ resolution (kHz) & 40 & 40 \\
Total time (mins) & 6 & 20 \\
\hline
\end{tabular}
\label{table:obs_params}
\end{table}
For both data sets, we use the same sky model, and simply switch in the \texttt{MS CLEAN} or shapelet model of Fornax A. The sky model is mostly based on the GLEAM catalogue~\citep{Hurley-Walker2017}, cross-matched to the following catalogues and frequencies: VLSSr, 74$\,$MHz~\citep{Lane2012}; TGSS ADR1, 150$\,$MHz~\citep{Intema2016}; MRC, 408$\,$MHz~\citep{Large1981}; SUMSS, 843$\,$MHz~\citep{Mauch2003}; NVSS, 1400$\,$MHz~\citep{Condon1998}. We use \texttt{PUMA}~\citep{Line2017} for the cross-match, using higher-resolution data for position or source morphology where possible. We use the sky model developed in~\citet{Procopio2017} in the EoR1 field. This sky model has been used by the MWA EoR analysis for a number of years, most recently in the limits published in~\citet{Trott2020}.

\begin{figure*}[htb!]
\centering
\includegraphics[width=0.9\textwidth]{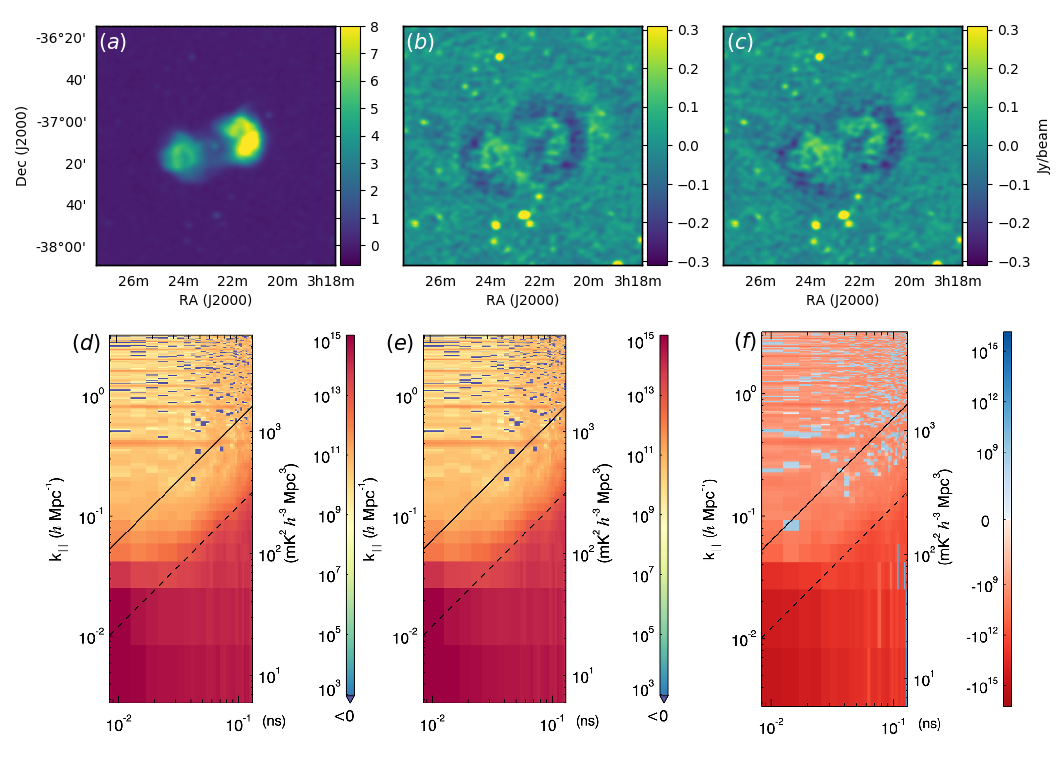}
\caption{Real data peel results for Fornax~A data detailed in Table~\ref{table:obs_params}: $(a)$ the calibrated data before subtraction; $(b)$ and $(c)$, the residuals after peeling the shapelet model and \texttt{MS CLEAN} model, respectively; $(d)$ and $(e)$, the 2D PS of the residuals after peeling the shapelet model and \texttt{MS CLEAN} model, respectively; $(f)$ the difference between the two 2D PS shown in $(d)$ and $(e)$, with the \texttt{MS CLEAN} PS subtracted from the shapelet PS. In $(e)$, red means the shapelet model subtracted more power during the peel, and blue means the \texttt{MS CLEAN} model subtracted more.}
\label{fig:real_peel_results}
\end{figure*}

We use the \texttt{RTS} to perform foreground subtraction in two steps: an initial direction-independent calibration to a sky model of 1000 sources; a subtraction step where a number of sources are fitted for first order ionospheric effects and peeled. Even though we are only comparing the Fornax~A models, we need to subtract off a number of sources in the sky to be able to see the difference in residuals, lest it be swamped by the rest of the sky. Typically, we peel off 1000 sources in the latter stage. As discussed in Section~\ref{sec:woden} however, the architecture of the \texttt{RTS} is currently such that creating a single calibrator source with thousands of components causes significant computational expense. We found as the Fornax~A data set had a higher time resolution, when processing with the \texttt{MS CLEAN} model, we could only peel off 500 sources before running into GPU memory issues. The processing time using the \texttt{MS CLEAN} model was also around 4 times slower than when using the shapelet model. The results of peeling 500 sources from the Fornax~A data are shown in Figure~\ref{fig:real_peel_results}, which shows both image residuals, and 2D PS.

In image space, the \texttt{MS CLEAN} and shapelet models leave similar residuals. A limitation of the shapelet fitting routine developed in this paper is exposed in Figure~\ref{fig:real_peel_results}b. The residuals are surrounded by a negative over-peel, which we attribute to sidelobe noise from the original image that was fitted into the shapelet model. This adds in a `floor' to the model, which the \texttt{RTS} then has to calibrate for, leading to a slight under-peel of the lobes, and a slight over-peel of the floor. Interestingly, a similar negative border surrounds the \texttt{MS CLEAN} residuals, which is possibly due to the largest scale fitted for during the \texttt{CLEAN}ing process, or a consequence of the peeling algorithms internal to the \texttt{RTS}. While the 2D difference power spectrum in Figure~\ref{fig:real_peel_results} suggests that the shapelet model has subtracted more power at all scales, the difference in the window is negligible, with the 1D PS looking near identical.

\begin{figure*}[htb]
\centering
\includegraphics[width=0.8\textwidth]{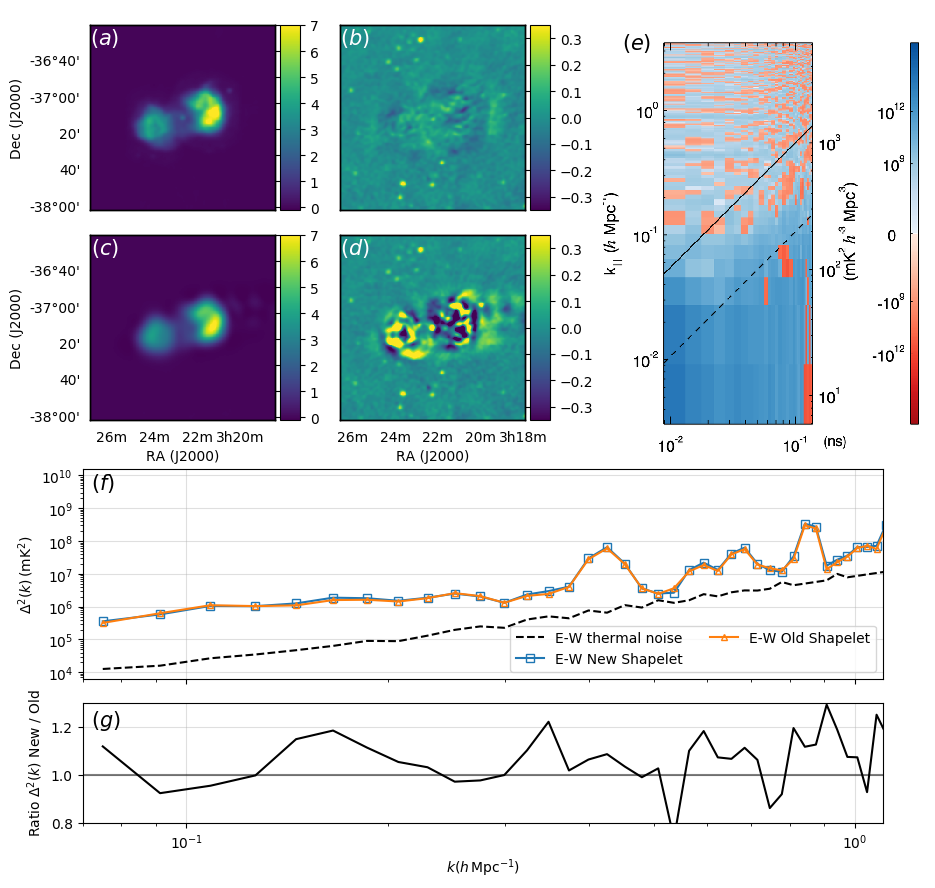}
\caption{Real data peel results for the EoR1 data detailed in Table~\ref{table:obs_params}: $(a)$ and $(b)$, the new phase I+II shapelet model simulated using the \texttt{RTS} and the residuals after peeling with this model, respectively; $(c)$ and $(d)$, the old shapelet model and residuals after peeling, respectively; $(e)$ an east-west polarisation difference 2D PS where the residuals of $(d)$ (old shapelets) are subtracted from $(b)$ (new shapelets); $(f)$ 1D PS of the residuals left behind after peeling the new shapelet mode (blue line with squares) and the old model (orange with triangles), with an estimate of the thermal noise in the east-west polarisation; $(g)$ the ratio of the 1D PS shown in $(f)$, with the new model divide by the old model. In $(f)$, red means the new model subtracted more power during the peel, and blue means the old model subtracted more. We only show east-west here, but the north-south display the same behaviour.}
\label{fig:real_peel_EoR1}
\end{figure*}

Given the small difference seen between the two models, we decided to compare our shapelet model to an earlier shapelet model created from early phase I data, for the EoR1 data. This model has historically been used by the Australian EoR team. Fornax~A has long been suspected of causing poor limits in the 1D PS. The results of comparing these two models are shown in Figure~\ref{fig:real_peel_EoR1}, and again we see little difference in the resultant 1D PS, regardless of the stark contrast in residuals left behind in the image. We explore possible reasons for the small differences in the following Section.

\section{Discussion}
\label{sec:discuss}
While the simulation results in Section~\ref{sec:sim_peel-results} indicate that the new shapelet model of Formax A we have generated should improve the 1D PS, the overall amplitude of the improvements seen in the simulated PS were several orders of magnitude lower than the current systematics present in real data. This is of course expected, given the simulations contained no other sources, were noiseless, and had no instrumental effects. We did expect however that this improvement might scale somewhat with real data, given the residuals seen in real data were far worse. The results using real data in Section~\ref{sec:real_data} resolutely disagree with this hypothesis, showing little to no difference.

\begin{figure}[htb]
\centering
\includegraphics[width=\columnwidth]{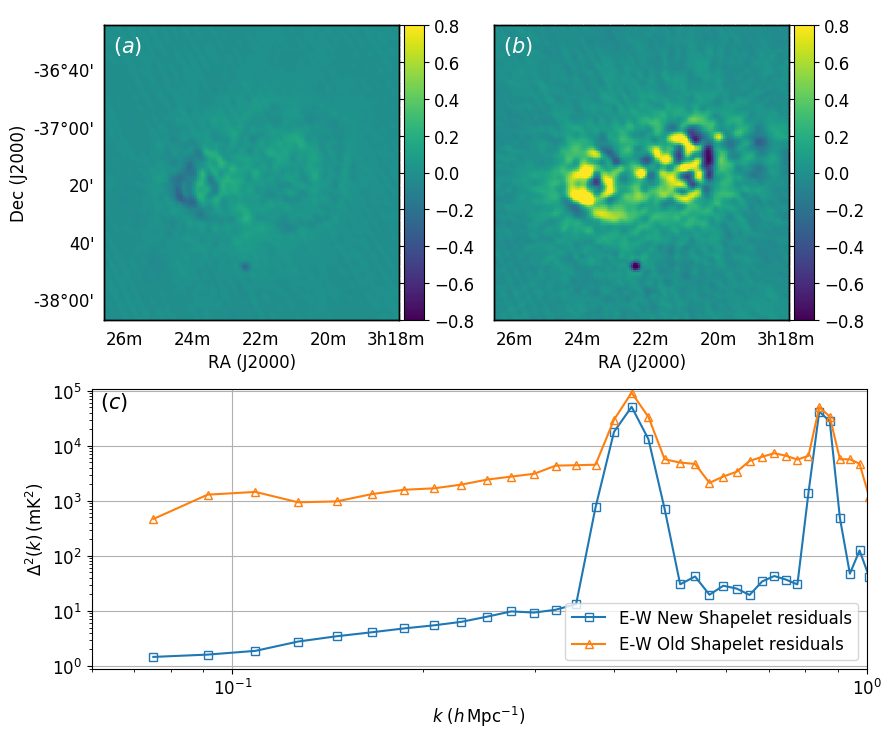}
\caption{Residual 1D PS comparing peeling 1000 simulated sources, including the new Fornax~A \texttt{MS CLEAN} model, when peeling sources including the new shapelet model (blue with squares) and including the old shapelet model (orange with triangles).}
\label{fig:rts-sim-1D}
\end{figure}

To address this, we return to simulations. Recently, functionality was added to the \texttt{RTS} to be run as a simulator, allowing the software to use an input sky model to generate visibilities, which it can then calibrate and peel. For a single 2~minute zenith observation, we weight our sky model by the MWA primary beam, and select the 1000 apparently brightest sources. We simulate these 1000 sources (with the primary beam and MWA bandpass), including the new \texttt{MS CLEAN} component model from the real image of Fornax~A. We then perform the standard direction-independent calibration, followed by direction-dependent ionospheric calibration to peel those 1000 sources, as we did with the real EoR1 data. We performed the peel with both the new and old shapelet models; this left us with two sets of residuals comparable to the those seen in Figure~\ref{fig:real_peel_EoR1}. We show the differences in the residual 1D PS in Figure~\ref{fig:rts-sim-1D}.

This shows a greater than two orders of magnitude difference in residuals after peeling with the two models, and also explains the lack of a difference in EoR1 data; the residuals in the \texttt{RTS} simulation vary of order $\sim 10^0$ - $10^4$, whereas in the real EoR1 data are of order $> 10^5$. Any improvement garnered through improving the Fornax~A model is seemingly blown away by other systematics in the flagging, calibration, and peeling of real EoR1 data. There are a number of potential systematics that could be masking the improvement from the new Fornax~A model. We list a handful of them here:
\begin{itemize}
    \item The calibration solutions from the \texttt{RTS} are not fit to be continuous in frequency. All twenty four 1.28$\,$MHz coarse channels are calibrated somewhat independently, which might inject false spectral structure into the residual visibilities.
    \item We use \texttt{AOFlagger}~\citep{Offringa2015} to flag RFI from our data. Early work suggests there may be a higher flagging occupancy seen in EoR1 data, when compared to data of the same frequency which is pointed at a different patch of sky.
    \item For each sky subtraction, Fornax~A is only one of 1000 sources being peeled. It is possible the rest of our sky model is inaccurate enough to leave residuals that cumulatively out-weigh those left behind after subtracting Fornax~A. Furthermore, there are the remaining sources that go unpeeled, that through instrumental and calibration errors, can still contribute power to the window.
\end{itemize}
Upcoming improved sky catalogues including LOBES~\citep[][in prep.]{Lynch2019} and GLEAM-X~\citep[][in prep.]{Hurley-Walker2019} which utilise phase II MWA data should help improve the sky model and subtraction to reduce residuals.

We also note here that in all simulations, Fornax~A was always given a single and known SI. In reality the SI varies across the object, giving the residuals in real data more spectral structure as well. This could be contributing to the lack of improvement seen in updating the model on real data. In our current implementation, the SI cannot be varied across the shapelet model, but could easily be controlled for each component of the \texttt{MS CLEAN} model. This spectral structure issue could possibly result in \texttt{MS CLEAN} components to perform better subtractions on sources with complicated spectral behaviour. Future work could investigate fitting shapelet models to multiple frequency bands, and fitting the derived coefficients $C_{p_1,p_2}$ as a function of frequency. This may allow spectral behaviour to vary with angular position in a shapelet source.

\section{Summary}
\label{sec:summarise}
We have successfully implemented a method to fit shapelets (\texttt{SHAMFI}) for use with the \texttt{RTS}, and explicitly specified how shapelets can be implemented to generate sky models in visibility space. We have written a new simulator, \texttt{WODEN}, in the \texttt{CUDA} language, allowing us to optimise simulating point sources, Gaussians, and shapelets in a consistent way. Through simulation, we see little difference in a 1D PS after peeling Fornax A, with shapelet and \texttt{MS CLEAN} models created from phase I simulated MWA observations. When adding in phase II (improved angular resolution) information to the simulation, we see that the shapelet model is able to subtract more power from the smallest $k$-modes than the equivalent \texttt{MS CLEAN} model. Interestingly, at small $k$-modes, the phase I+II \texttt{MS CLEAN} model subtracts less power than the \texttt{MS CLEAN} model made from just phase I data. Adding in the smaller angular resolution information does however improve the peel at large $k$-modes. We suggest that the change in scale of the synthesized PSF could be biasing the \texttt{MS CLEAN} components to smaller angular scales, but further work is required to verify this. Through simulation, we are also find that with truncation, we are able to still better subtract power using the phase I+II shapelet model for the same computational cost. 

No difference could be measured with a 1D PS between peeling with a shapelet or \texttt{MS CLEAN} model, when using the \texttt{RTS} and \texttt{CHIPS}, on real data. Simulations suggest that modest, but not insignificant, improved residuals at the smallest $k$-modes could be achieved by using shapelets over \texttt{MS CLEAN} components, when using both phase I+II data. Further simulations show these improvements are orders of magnitude smaller than current systematics in our analysis and data. Pragmatically, when using the current Australian EoR pipeline, a model of Fornax A is essential, as a \texttt{MS CLEAN} model takes 4 times more CPU hours.

A potential weakness of \texttt{SHAMFI} was highlighted in Section~\ref{sec:real_data}, where it was found shapelets tend to fit the sidelobe noise of other sources surrounding the target object. One could avoid this by fitting shapelets just to the \texttt{MS CLEAN} components, rather than the restored image. In doing this however, one will miss any diffuse emission that hasn't been deconvolved, limiting the improvement shapelets can bring over \texttt{MS CLEAN} components.

In this paper, we focus on a single object, Fornax A. In reality, thousands of sources must be peeled, ranging from a simple to complicated morphology. Shapelets are well suited to modelling sources with extended diffuse emission. These types of sources are well observed by the MWA, due to the compact layout of core receiving elements giving the MWA an excellent surface brightness sensitivity. Sources such as NGC 253~\citep{Kapinska2017} and extended sources described in~\citet{Procopio2017}, which lie in MWA foregrounds, could be accurately modelled with shapelets.

Although not able to immediately make an impact on EoR limits, the simulations suggest that as the MWA EoR analysis progresses in pushing down the systematics from the instrument and software, shapelet models of Fornax~A should improve the limits found. With an eye on future instruments such as the upcoming SKA\_{}LOW, with far greater angular resolution, we note that the number of \texttt{MS CLEAN} components needed to create models will also increase. With the excellent response to truncation shown by shapelets, they might be a vital method to include highly complex sources in an EoR foreground sky model, for a lower computational cost.

\section*{Acknowledgements}
This scientific work makes use of the Murchison Radio-astronomy Observatory, operated by CSIRO. We acknowledge the Wajarri Yamatji people as the traditional owners of the Observatory site. Support for the operation of the MWA is provided by the Australian Government (NCRIS), under a contract to Curtin University administered by Astronomy Australia Limited. We acknowledge the Pawsey Supercomputing Centre which is supported by the Western Australian and Australian Governments.

This research was supported by the Australian Research Council Centre of Excellence for All Sky Astrophysics in 3 Dimensions (ASTRO 3D), through project number CE170100013. CMT is supported by an ARC Future Fellowship under grant FT180100321.

This work was supported by resources awarded under Astronomy Australia Ltd's merit allocation scheme on the OzSTAR national facility at Swinburne University of Technology. OzSTAR is funded by Swinburne University of Technology and the National Collaborative Research Infrastructure Strategy (NCRIS).

During this work we made extension use of the \texttt{kvis}~\citep{Gooch1996} and \texttt{DS9}~\citep{Joye2003} FITS file image viewers, and the \texttt{astropy}~\citep{astropy2013, astropy2018} and \texttt{aplpy}~\citep{aplpy2012,aplpy2019} \texttt{python} modules.

We thank the anonymous referee for their useful and constructive comments.
\bibliographystyle{pasa-mnras}
\bibliography{reduced_refs}

\appendix
\section{WSClean command}
\begin{figure}[h]
    \begin{verbatim}
   
###########################################
wsclean -size 512 512 -niter 100000 \
  -name ./images/woden_VLA-ForA_phase1+2 \
  -auto-threshold 0.5 -auto-mask 3 \
  -pol I -multiscale -weight uniform \
  -scale 0.004 -j 12 -mgain 0.85 \
  -no-update-model-required \
  -save-source-list \
  -channels-out 4 -join-channels \
  -fit-spectral-pol 1 \
  ./data/*.ms
###########################################
  \end{verbatim}
    
    \caption{Example \texttt{WSClean} command used on simulated data}
    \label{code:wsclean_command}
\end{figure}

\section{Software versions}
\label{app:soft-versions}

\begin{table}[h]
\centering
\renewcommand{\arraystretch}{1.5}
\caption{Versions of software used}
\begin{tabularx}{.99\columnwidth}{>{\setlength\hsize{.37\hsize}} X | >{\setlength\hsize{0.63\hsize}} X}
\hline
\textbf{Software} & \texttt{git -{}-describe} \\
\hline
\hline
WSClean & wsclean2.8-37-gda89428 \\
\hline
IDG & 0.6-178-g5736086c \\
\hline
WODEN (VLA simulations) & 1d1da63 \\
\hline
WODEN (all other simulations) & 854d9c8 \\
\hline
SHAMFI & 1b81779 \\
\hline
CASA & casa-pipeline-release-5.6.2-2.el7 \\
\hline
CHIPS & 71b251d \\
\hline
RTS & 3f3b3210 \\
\hline
\end{tabularx}
\label{table:software}
\end{table}

\end{document}